\documentclass[prl, aps, twocolumn, superscriptaddress,noshowpacs]{revtex4}
\usepackage[english]{babel} 
\usepackage[english]{layout}
\usepackage{amsmath,amsfonts,amssymb}
\usepackage{graphicx}
\usepackage{float}
\usepackage{color}
\usepackage{braket}
\usepackage{version}
\usepackage{array,multirow,makecell}
\newcommand{\sign}{\text{sign}}

\begin{document}

\title{Parametric amplification of topological interface states in synthetic Andreev bands}
\author{I. Septembre}
\affiliation{Institut Pascal, PHOTON-N2, Universit\'e Clermont Auvergne, CNRS, SIGMA Clermont, F-63000 Clermont-Ferrand, France}
\author{S. Koniakhin}
\affiliation{Institut Pascal, PHOTON-N2, Universit\'e Clermont Auvergne, CNRS, SIGMA Clermont, F-63000 Clermont-Ferrand, France}
\affiliation{St. Petersburg Academic University - Nanotechnology Research and Education Centre of the Russian Academy of Sciences, 194021 St. Petersburg, Russia}
\author{J. S. Meyer}
\affiliation{Univ. Grenoble Alpes, CEA, IRIG-Pheliqs, F-38000 Grenoble, France}
\author{D. D. Solnyshkov}
\affiliation{Institut Pascal, PHOTON-N2, Universit\'e Clermont Auvergne, CNRS, SIGMA Clermont, F-63000 Clermont-Ferrand, France}
\affiliation{Institut Universitaire de France (IUF), 75231 Paris, France}
\author{G. Malpuech}
\affiliation{Institut Pascal, PHOTON-N2, Universit\'e Clermont Auvergne, CNRS, SIGMA Clermont, F-63000 Clermont-Ferrand, France}

\begin{abstract}
A driven-dissipative nonlinear photonic system (e.g.  exciton-polaritons) can operate in a gapped  superfluid regime. We theoretically demonstrate that the reflection of a linear wave on this superfluid is an analogue of the Andreev reflection of an electron on a superconductor. A normal region surrounded by two superfluids is found to host Andreev-like bound states. These bound states form topological synthetic bands versus the phase difference between the two superfluids. Changing the width of the normal region allows to invert the band topology and to create ``interface'' states. Instead of demonstrating a linear crossing,  synthetic bands are attracted by the non-linear non-Hermitian coupling of bosonic systems which gives rise to a self-amplified strongly occupied topological state.  
\end{abstract}

\maketitle
Topological physics relies on the specific structure of the eigenstates of Hamiltonians in a parameter space. It became one of the most active fields of research of the last decades. Topological invariants were successively used to characterise superfluid excitations (quantum vortices) \cite{Onsager1949,Pitaevskii}, solitons in polyacethylene \cite{Su1979},  Landau levels \cite{thouless1982quantized}, electronic Bloch bands in solids \cite{haldane1988model}, and more generally energy bands in periodic media \cite{haldane2008possible}. A related concept is the bulk-boundary correspondence \cite{Hatsugai1993,Mong2011}, which associates the change of the band topology with the gap closing and therefore with the existence of an interface state between media of different topologies. Such interface states can demonstrate unique properties, such as the chiral (one-way)  transport in topological insulators \cite{Hasan2010}.

The recent proposals and implementations of topological media supporting protected edge states in non-linear systems opened new possibilities \cite{Lumer2013,Bardyn2016,bleu2016interacting,Bleu2017x,Gulevich2017,Kartashov2017,bleu2018robust,kruk2019nonlinear,Smirnova2020}. The most well-known examples are the so-called topological lasers, where lasing occurs at the topological edge states of a photonic lattice \cite{solnyshkov2016kibble,st2017lasing,bahari2017nonreciprocal,bandres2018topological}.
Another recent orientation is synthetic topological matter, where the parameters of the Hamiltonian are not imposed by the system itself (such as the wave vector in Bloch bands), but rather externally, by experimental conditions. This approach allows to explore regimes which cannot be accessed otherwise, such as the 4D quantum Hall effect \cite{lohse2018exploring}. 
One possibility to implement synthetic topological matter relies on using Josephson junctions between conventional superconductors  \cite{riwar2016multi}. The dependence of Andreev bound states versus the phase difference between the superconductors which form the junction allows to define synthetic bands with the synthetic dimensionality controlled by the number of terminals. These bands were found to be topological, showing Weyl singularities, and significant efforts are made to implement them \cite{draelos2019supercurrent,Pankratova2020}. Andreev reflection occurs at the interface of a superconducting medium \cite{andreev1964thermal}, where an incoming electronic wave undergoes an anomalous reflection, where its wavevector is reversed and is interpreted as a hole excitation associated with opposite charge and mass. The analogy between Andreev reflection and the reflection of a wave over a Bose-Einstein Condensate has been studied theoretically  \cite{Zapata2009}. In photonic systems, one can implement a superconductor analog using resonant driving,  typical for cavity exciton-polaritons \cite{Carusotto2013,kavokin2017microcavities}. In such a case, a gap opens in the spectrum of elementary excitations of the pumped modes which corresponds to the formation of a ``gapped superfluid''. It is, moreover, possible to modulate  the pump intensity in space  to realize two or more gapped regions with controllable phase,  separated by normal (non-superfluid) regions \cite{goblot2016phase,Koniakhin2019,Claude2020}. 

In this work, we study the analog of the Andreev reflection of an incident wave on a gapped driven-dissipative polariton superfluid. We show the existence of Andreev-like bound states between the two driven superfluids. The topology of the synthetic bands defined versus the phase difference between the two superfluids is found to be non-trivial. The interface states between the regions of inverted topology are found to be self-amplified. They are macroscopically-occupied topologically protected Andreev-like states.

\emph{The model.} We consider a resonantly-pumped 1D non-linear optical media described by the driven-dissipative Gross-Pitaevskii equation for scalar particles, which is similar to the Lugiato-Lefever equation \cite{Lugiato1987}: 
\begin{equation}\label{GP}
    i\hbar \frac{\partial\psi}{\partial t}=\left[ -\frac{\hbar^2}{2m}\frac{\partial^2}{\partial x^2} -i\gamma +\alpha |\psi|^2 \right] \psi +P,
\end{equation}
 where $P=e^{-i \omega_p t}$ is the quasi-resonant pumping term at normal incidence, $\alpha>0$ is the repulsive interaction term, and $\gamma$ the decay rate. The pumped mode dispersion is parabolic (mass  $m$),  typical for photons or exciton-polaritons in a microcavity. All polarisation effects are neglected. The spatially-homogeneous solution $\psi_s e^{-i\omega_p t}$ shows a bistable behaviour \cite{Baas2004} with a large occupation of the pumped mode above the bistable threshold. To simplify the analytics, we consider the limit $\gamma\to 0$ (see \cite{suppl}). The wavefunction describing  weak excitations of the macro-occupied mode reads:
\begin{equation}\label{wpwf}
    \psi(x,t)=e^{-i\omega_p t}\left(\psi_s+ u e^{i k x}e^{-i\omega t}+v^* e^{-i k x} e^{i\omega^* t}\right).
\end{equation}
Here, one plane-wave excitation is coupled by the non-linear term to its complex conjugate. Inserting this wavefunction into Eq. \eqref{GP} allows one to derive the Bogoliubov-de Gennes equations:
\begin{equation}\label{BdG0}
\begin{pmatrix}
 \mathcal{L} && \alpha \psi_s^2 \\
 -\alpha \psi_s^{*2}  && -\mathcal{L^*}
\end{pmatrix}\begin{pmatrix}
u\\v
\end{pmatrix}=E\begin{pmatrix}
u\\v
\end{pmatrix},
\end{equation}
where $\mathcal{L}=\left(\epsilon_k-E_p+2\alpha n\right)$. The coefficients $u,~v$ are the Bogoliubov coefficients, $\epsilon_k=\hbar ^2 k ^2/2m$ and $\psi_s=\sqrt{n}e^{i\phi}$.
The energy $E=\hbar\omega$, measured with respect to the laser detuning $E_p=\hbar\omega_p$, is found by cancelling the determinant and reads:
\begin{equation}\label{E(k)}
    E^2=\left( \epsilon_k+\alpha n -E_p\right)\left( \epsilon_k+ 3\alpha n -E_p\right).
\end{equation}

A bogolon is formed by the superposition of two plane waves of amplitude $u$ and $v^*$ at energies $E,~ -E$. Considering $E$ positive, the ratio $|u|/|v|$ is fixed by Eq.~\eqref{BdG0} and is larger than 1. The specific choice of a normalisation condition $|u|^2-|v|^2 = 1$ allows to define a bogolon as a particle of positive total energy $(|u|^2-|v|^2)E$, but containing amplitudes at both $E$ and $-E$.
When $\alpha n>E_p$, the spectrum of allowed energies (containing both positive and negative parts) shows a gap of magnitude $2\Delta$ centered on the pump energy: 
\begin{equation}
    \Delta=\sqrt{\left(\alpha n -E_p\right)\left(3\alpha n-E_p \right)}.
\end{equation}

\begin{figure}
    \centering
    \includegraphics[width=7cm]{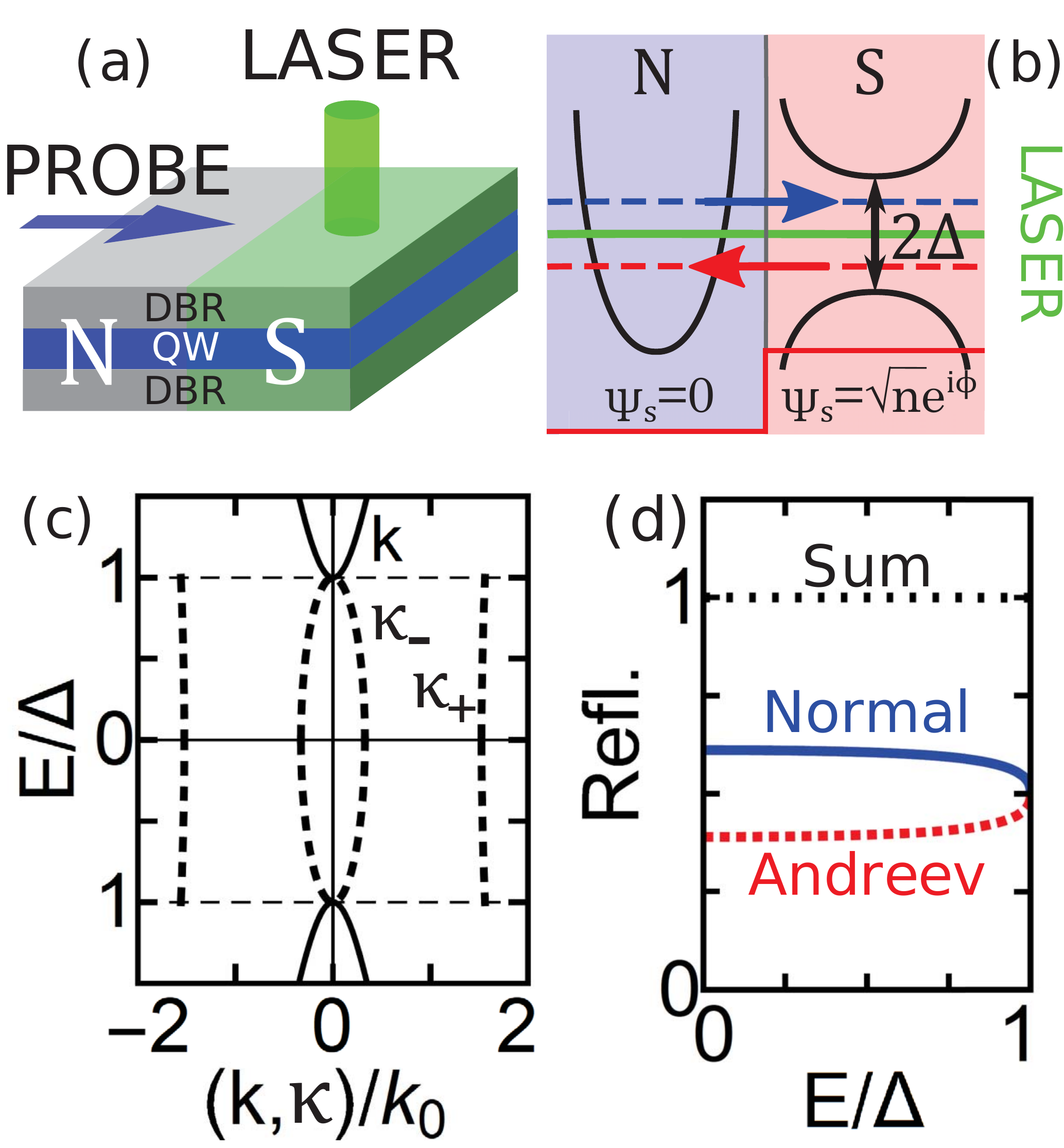}
    \caption{(a) Scheme of the system considered: a microcavity with a quantum well (QW)  and distributive Bragg reflectors (DBR). A half-space (S) is pumped by a laser under normal incidence, while the other half (N) is not. The probe comes from the normal region towards the driven one. (b) Scheme of the energy spectrum and order parameter in each region. (c) Bogolon energy versus the wavevector $k$ (propagative states, solid lines) and the inverse decay lengths $\kappa_\pm$ (evanescent states, dashed lines). $k_0=k|_{E=0}=\sqrt{2mE_p}/\hbar$. (d) Normal and Andreev reflection coefficients versus energy.}
    \label{fig1}
\end{figure}

Next, we consider two semi-infinite regions (Fig.~\ref{fig1}(a)). The left part is not pumped and characterized by a parabolic dispersion. The right part is resonantly pumped and described by the 1D GP equation. We consider an incident wave coming from the normal part to the pumped area at an energy within the gap. We have therefore to consider evanescent bogolon modes, instead of the usual propagative ones. The corresponding wavefunction reads:
\begin{equation}
    \psi(x,t)=e^{-i\omega_p t}\left(\psi_s+ u e^{-\kappa x}e^{-i\omega t}+v^* e^{-\kappa x} e^{+i\omega^* t}\right).
\end{equation}

The Bogoliubov-de Gennes equations read as Eq.~\eqref{BdG0}, but with $\mathcal{L}=\left(-\epsilon_{\kappa}-E_p+2\alpha n\right)$ and $\epsilon_k=\hbar ^2 {\kappa}^2/2m$. The characteristic equation provides conditions on the inverse decay length values:
\begin{equation}
    \kappa_\pm = \sqrt{2m \left(2\alpha n-E_p\pm\sqrt{(\alpha n)^2+E^2} \right)} \bigg/ \hbar.
\end{equation}

Importantly, at a given positive $E$, there exist two different evanescent waves with two different inverse decay lengths and two different eigenvectors: 
\begin{equation}
    \begin{array}{r c l}\label{u-v}
      u_\pm & = & \frac{\sqrt{\sqrt{(\alpha n)^2+E^2}\mp E}}{\sqrt{2E}}e^{i\phi},\\
      v_\pm & = &\pm \frac{\sqrt{\sqrt{(\alpha n)^2+E^2}\pm E}}{\sqrt{2E}}e^{-i\phi}.
   \end{array}
\end{equation}
The ``minus'' state has a longer decay length and $|u_-|>|v_-|$. Its dominant component has an energy $E>0$ and the state is normalized as $|u_-|^2-|v_-|^2=1$. It is quite similar to propagative bogolons, continuing their dispersion within the gap, as shown on Fig.~\ref{fig1}(c). We call this type of state, where the positive energy component is larger, a ``particle''. The ``plus'' state has a shorter decay length and $|u_+|<|v_+|$. Its dominant component has an energy $-E<0$ as $|u_+|^2-|v_+|^2=-1$. This evanescent solution has no propagative counterpart \cite{suppl}. The $\kappa_+$ branch shown in Fig.~\ref{fig1}(c) is disconnected from the propagative states dispersion.  This type of solution never described before to our knowledge can be assimilated to a ``hole'' state. It is associated with a local decrease of the particle density with respect to the homogeneous superfluid. It plays a crucial role in the Andreev-like reflection, as described below.

As shown in Fig.~\ref{fig1}(b), we consider a plane wave of energy $0<E<\Delta$  incident on the pumped region. It excites the two above-mentioned types of evanescent bogolons, which provoke reflection at the energies $E$ and $-E$, respectively.
The wavefunction in the normal (left) region reads:
\begin{equation}\label{psiN}
    \psi_N=\begin{pmatrix} Ae^{ik_1x} + Be^{-ik_1x}\\ 
    Ce^{ik_2x} + De^{-ik_2x}\end{pmatrix},
\end{equation}
with $k_{1,2}~=~\sqrt{2m\left ( E_p\pm E \right )} / \hbar$, valid for $\Delta<E_p$ (see \cite{suppl} for other cases with evanescent states in the normal region). In the superfluid, the wavefunction combines two evanescent waves:
\begin{equation}
   \psi^S= \eta_+ \begin{pmatrix}u_+ \\ v_+ \end{pmatrix}  e^{-\kappa_+ x} +\eta_- \begin{pmatrix}u_- \\ v_-\end{pmatrix}  e^{-\kappa_- x}.\label{psiS}
\end{equation}

To compute the reflection coefficients, we take $A=1/\sqrt{v_1}$, $C=0$. In that case $B=r_{Np}/\sqrt{v_1}$,  $D=r_{Ap}/\sqrt{v_2}$ are the normal and Andreev reflection coefficients for an incident ``particle'' (dominant positive energy component). The group velocities allow to conserve the current. Similarly, $A=0$, $C=1/\sqrt{v_2}$ corresponds to $B=r_{Ah}/\sqrt{v_1}$,  $D=r_{Nh}/\sqrt{v_2}$, the reflection coefficients for an incident "hole"  having a  dominant negative energy component. The continuity of the wavefunctions and of their derivatives at the interface gives an analytical expression for these reflection coefficients (see \cite{suppl}) which are plotted in Fig.~\ref{fig1}(c) for parameters  $m=5\times 10^{-5}m_0$, $E_p=0.5$~meV, and $\alpha n \approx 1.11 E_p$, characteristic for GaAs exciton-polaritons. This yields a gap $\Delta \approx 0.5 E_p$. The Andreev-like reflection is comparable in amplitude to the normal reflection. Ultimately, the phenomenon occurring here is very similar to the Andreev reflection, but can also be interpreted as a non-linear frequency conversion. An incoming wave at the frequency $\omega_p \pm \delta \omega$ is partially reflected both at $\omega_p \pm \delta \omega$ and at $\omega_p \mp \delta \omega$. In the case of in-gap energies, the reflection (normal and Andreev together) is total, since $|r_{N{p,h}}|^2+|r_{A{p,h}}|^2=1$. Optical phase conjugation, discovered and studied in the 70's in nonlinear optics  \cite{yariv1978phase} also shows a strong analogy with Andreev reflection \cite{van1991andreev,PhysRevA.56.4216}.

\begin{figure}
    \centering
    \includegraphics[width=7.5cm]{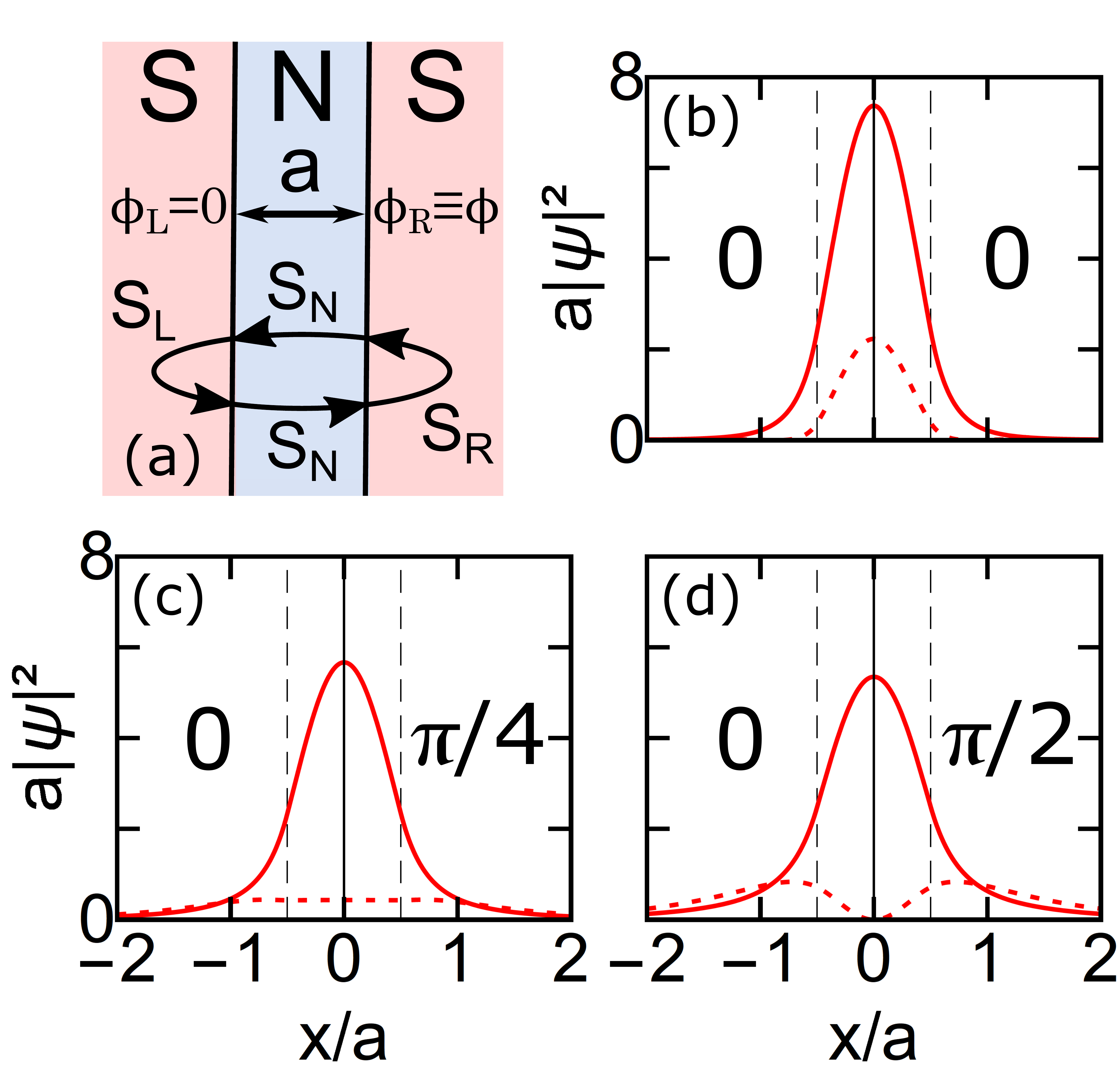}
    \caption{(a) Schematic representation of the bosonic SNS junctions with its principal characteristics. The matrices depicting each scattering event are also represented. (b)-(d) Probability density at each energy for $a=2.6\,\mu$m and a phase difference of (b) $0$, (c) $~\pi/4$ and (d) $\pi/2$.}
    \label{fig2}
\end{figure}

\emph{Andreev bound states analog.}
The next step is to consider a Superfluid - Normal - Superfluid (SNS)  junction (Fig.~\ref{fig2}(a)). The only difference between the two superfluids is their phase, $\phi_L=0$ and $\phi_R \equiv \phi$, respectively. The width of the normal region is $a$. This structure exhibits trapped states (similar to quantum well eigenstates), determined by the density profile even without the Andreev reflection. The latter provides a correction and generates a second energy component (bogolon image) for each of these states.

The energy components of the Andreev bound states are computed using the scattering matrix formalism. The scattering matrices for the reflection phenomena at each interface read:
\begin{equation}
    S_{R(L)}=\begin{pmatrix}
    r_{Np}^{R(L)} & r_{Ah}^{R(L)} \\
    r_{Ap}^{R(L)} & r_{Nh}^{R(L)}
\end{pmatrix},
\end{equation}
where $S_N$ describes the propagation in the normal region \cite{suppl}. The total scattering matrix reads $S_T=S_LS_{N}S_RS_{N}$ [see Fig.~\ref{fig2} (a)].
A bound eigenstate exists, if its energy components $\pm E$ satisfy the condition~\cite{beenakker1991universal}:
\begin{equation}\label{eqdet}
    \det[\mathbb{I}-S_T]=0.
\end{equation}

The eigenvectors of $S_T$ determine the wavefunction via Eqs.~\eqref{psiN},\eqref{psiS} and allow one to determine if a state is either particle-like or hole-like. Depending on parameters, the eigenenergies can be either real (stationary Andreev-like bound states)  or imaginary (self-amplified bound states).  Figures~\ref{fig2}(b) and (c) show two examples of hole-like bound states. In Fig.~\ref{fig2}(b), $\phi=0$ and both energy components have the same parity ($s$-like state). In Fig.~\ref{fig2}(c), $\phi=\pi/2$. The main component keeps its parity, whereas the bogolon image becomes $p$-like, because of the phase shift. We compute the probability current $J_\pm$ in the normal region \cite{suppl}, similar to the Josephson current in superconducting junctions:
\begin{equation}
    J_\pm \approx \pm J_0  \sin{2\phi}.
\end{equation}
The total probability current of one component at the energy $+E$ is fully compensated by the current at the energy $-E$. 

 \begin{figure}
    \centering
    \includegraphics[width=7.5cm]{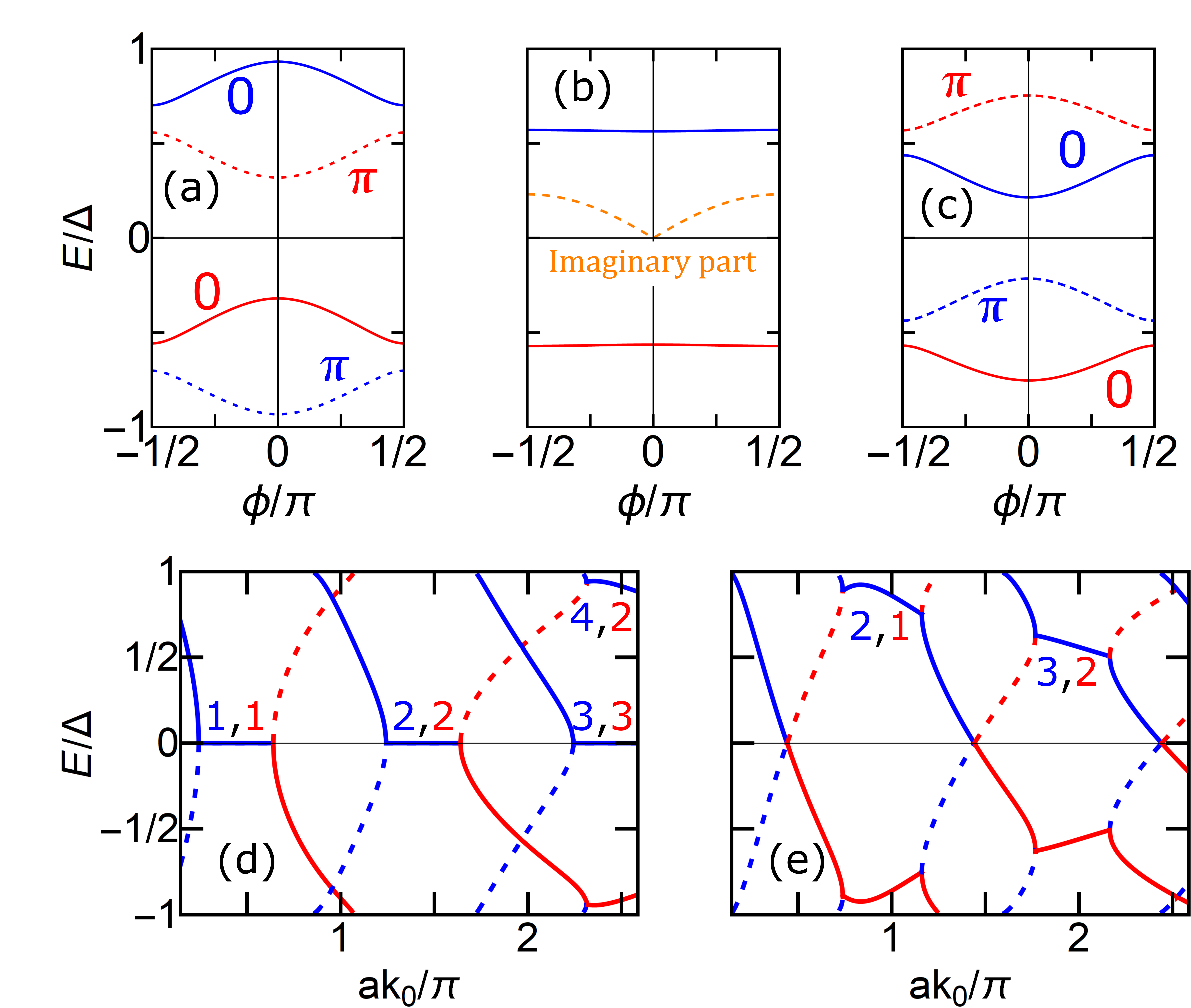}
    \caption{Synthetic bands and their Zak phase (a) ``before'' the instability ($a=6.8\,\mu$m), (b) at the maximum of the instability ($a=7.6\,\mu$m), (c) ``after'' the instability ($a=8.4\,\mu$m). (d,e) Bound state energies versus $a$ for $\phi=0$ (d) and $\phi=\pi/2$ (e) Blue lines: particle-like states. Red lines: hole-like states; the dominant (minority) component is in solid (dashed) line.  The pairs of numbers indicate the states from which the crossing bands take their origin.}
    \label{fig3}
\end{figure}

 \begin{figure}
    \centering
    \includegraphics[width=7.5cm]{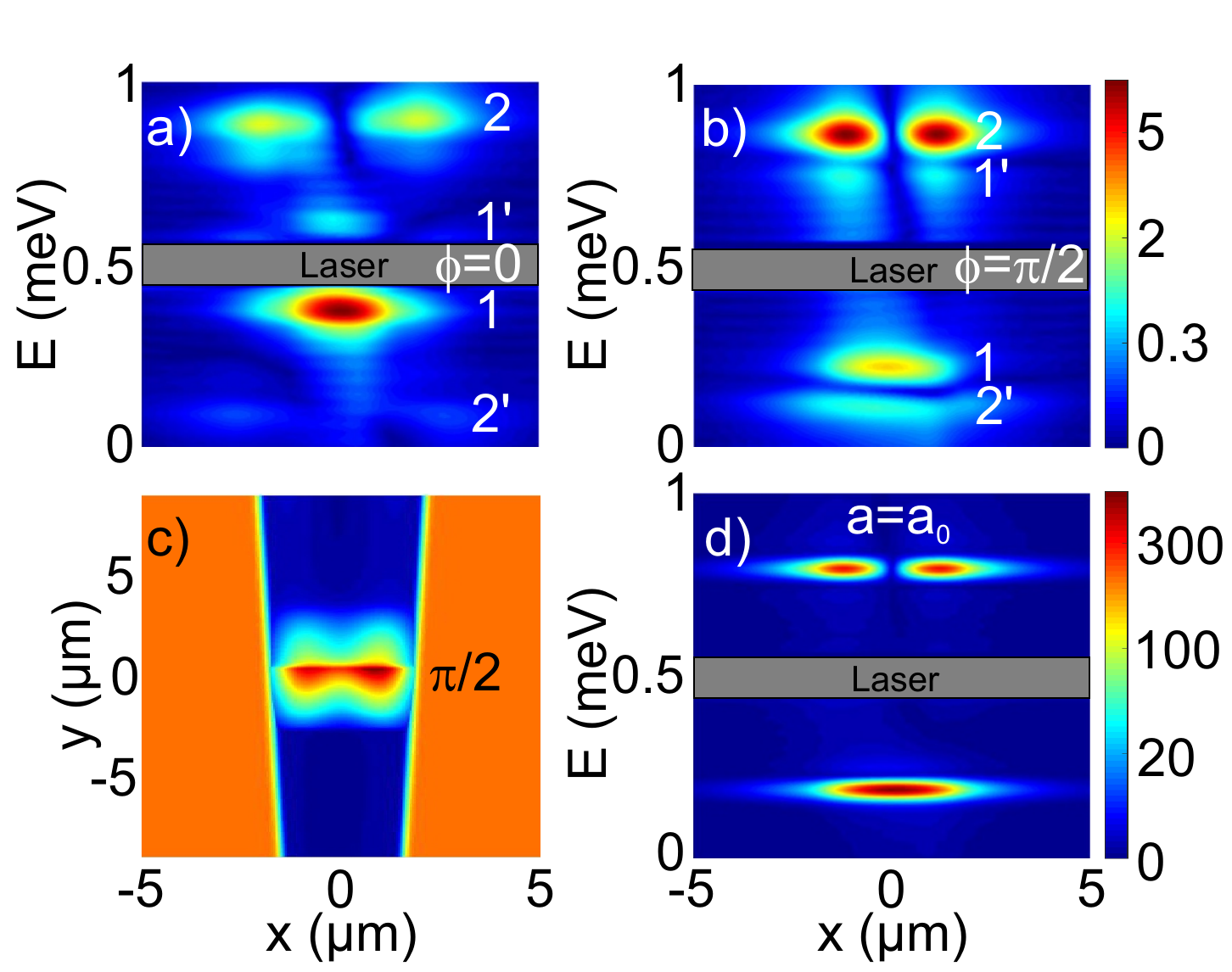}
    \caption{Numerically calculated spectrum of the Andreev states. Two energy components are visible for each of the two states (marked $1,1'$ and $2,2'$) for $a<a_0$: (a) $\phi=0$, (b) $\phi=\pi/2$. (c) Numerically calculated emission intensity for a system with varying width $a$. The amplified OPO states are visible at $y=0$. (d) The spectrum of the resonantly amplified states at $y=0$. The laser energy is cut out for all images.}
    \label{fig4}
\end{figure}

\emph{Topological synthetic bands.}
Both positive and negative energy components of a bound state form synthetic energy bands with respect to $\phi$. For a given energy band, the wavefunction is a superposition  of two counter-propagating plane waves, as defined by Eq.~\eqref{psiN}. The amplitudes of these two plane waves define a  pseudospinor $X^+~\sim~(A,B)^T$ for positive energies and $X^-\sim (C,D)^T$ for negative energies. The evolution of these pseudospinors along the band allows to compute the Zak phase 
\cite{Zak1989,Delplace2011}:
 \begin{equation}
 \Phi_{Zak}^\pm  = \int {\left\langle {X^\pm}
  \mathrel{\left | {\vphantom {X {i\frac{{\partial X}}{{\partial \varphi }}}}}
 \right. \kern-\nulldelimiterspace}
  {{i\frac{{\partial X^\pm}}{{\partial \phi }}}} \right\rangle } d \phi.
 \end{equation}
 
 Figure~\ref{fig3}(a-c) shows the synthetic bands and their Zak phases for three  thicknesses $a$. In Fig.~\ref{fig3}(a,c) all energies are real. The bands shown in blue correspond to a particle-like state with a dominant positive energy part (solid lines). Its negative energy counterpart (bogolon image) is smaller in amplitude (dashed lines) and shows a non-zero Zak phase. Indeed, at $\phi=0$, $C=\pm D$. The current is zero, and the associated pseudospin lies in the plane of a Bloch sphere representation. At $\phi\approx \pi/4$, $D=0$. The current is maximal, and the pseudospin points towards the pole. Finally, at  $\phi=\pi/2$, $C=\mp D$: the symmetry of the state has changed. Between $\phi=-\pi/2$ and $\phi=\pi/2$, the pseudospin covers a full great circle of the Bloch sphere (constrained by the mirror symmetry of the problem, see \cite{suppl}), and the accumulated Zak phase is $\pi$. On the other hand, the pseudospin of the majority component slightly moves towards the pole at $\phi\approx \pi/4$ to go back to its original position at $\phi=\pi/2$. In general, the band associated with the dominant energy component of the bogolon (the original trapped state) shows a null Zak phase, whereas the  minority component (the bogolon image) is topologically non-trivial.
 
Fig.~\ref{fig3}(a,c) show the band topology inversion between two values of $a$ which implies
the gap closing for a critical thickness $a_0$. This topological band crossing normally gives rise to Dirac, or Weyl points, depending on the system's dimensionality. It is also at the heart of topologically protected edge or interface states. In our bosonic system of interacting particles, crossing bands interact through a non-linear non-Hermitian coupling [Eq.~\eqref{BdG0}]. Fig.~\ref{fig3}(b) shows the states at the critical thickness $a_0$, where the gaps are closing. Instead of simply crossing, the bands merge: their real parts become flat, while the imaginary parts are opposite. The positive imaginary part means that the topological state corresponding to the crossing is amplified and becomes strongly populated. This amplification occurs when the two bogolon components are resonant with the linear eigenstates of the potential trap. Fig.~\ref{fig3} (d) and (e) show for $\phi=\pi/2$ and $\phi=0$,  respectively, the mode energy versus $a$. For $\phi=0$, amplification occurs when the majority and minority component have the same parity [$(n,n)$], when the potential well states $n=1,2,\ldots$ are resonant with the laser.  For $\phi=\pi/2$, the amplification occurs when the bogolon image of a state of given parity becomes resonant with an original trapped state of different parity.

\emph{Numerical simulations.} To confirm the analytical theory, we perform numerical simulations, solving Eq.~\eqref{GP} over time with a weak probe exciting the bogolon states, and a finite lifetime $\tau=\hbar/2\gamma=30$~ps. The width $a$ varies versus $y$ as: $a(y)=a_0+y/l$, where $l=64$~$\mu$m. The calculated spectra are presented in Fig.~\ref{fig4} for  $a<a_0$ for $\phi=0$ (a) and $\phi=\pi/2$ (b). The two energy components of each of the Andreev states are marked in the Figure (e.g. $1$ is the ``original'' trapped state and $1'$ is its bogolon image). The change of the symmetry of the bogolon images $1'$ and $2'$ is clearly visible. The gap closes at $a=a_0$, where each original state is resonant with the bogolon image of the other: $E_{2}=E_{1'}$ and $E_{1}=E_{2'}$, and their symmetries also coincide for $\phi=\pi/2$. As a result, the Andreev states are amplified and dominate the spectrum  (Fig.~\ref{fig4}(d)).  Figure~\ref{fig4}(c) shows the spatial distribution of emission, with the amplified Andreev state  visible at $y=0$.
We stress that the normal part of the junction is a \emph{non-interacting} medium, otherwise solitons are formed there  \cite{goblot2016phase,Koniakhin2019,Claude2020}, and the approximations used to make our calculations are not valid anymore.

To summarize, we predict an analogue of the Andreev reflection in a photonic driven-dissipative gapped superfluid. Such systems can be used to form bosonic SNS junctions hosting Andreev-like bound states. These bound states form topologically nontrivial synthetic bands as a function of the phase difference between the pumping lasers. By changing the width of the normal region, one can invert their topology, and the associated topologically protected interface states are found to be self-amplified, giving rise to strongly emitting topologically protected photonic modes.

\begin{acknowledgments}
We thank O. Bleu for useful discussions. We acknowledge the support of the projects EU "QUANTOPOL" (846353), "Quantum Fluids of Light"  (ANR-16-CE30-0021), of the ANR Labex GaNEXT (ANR-11-LABX-0014), and of the ANR program "Investissements d'Avenir" through the IDEX-ISITE initiative 16-IDEX-0001 (CAP 20-25). J. S. Meyer thanks the project "Hybrid" (ANR-17-PIRE-0001).
\end{acknowledgments}

\bibliography{biblio} 

\renewcommand{\thefigure}{S\arabic{figure}}
\setcounter{figure}{0}
\renewcommand{\theequation}{S\arabic{equation}}
\setcounter{equation}{0}

\section{Supplemental Materials}

\subsection{Propagative and evansecent bogolons}
In the main text, the spectrum of elementary excitations of bogolons is proven to present a gap. This gap directly comes from the dispersion relation:
\begin{equation}
    E^2=\left( \epsilon_k+\alpha n -E_p\right)\left( \epsilon_k+ 3\alpha n -E_p\right).
    \label{propdisp}
\end{equation}
At first glance, in the propagative case, this provides two solutions for the norm of the wavevector $|k|$ (that is, four solutions for $k$):
\begin{equation}
    |k_\pm| = \sqrt{2m \left(E_p - 2\alpha n \pm\sqrt{(\alpha n)^2+E^2} \right)} \bigg/ \hbar.
\end{equation}
However, the solution $k_-$ is actually imaginary, since in the propagative case $E>\Delta$, which yields $\sqrt{(\alpha n)^2+E^2} > 2 \alpha n - E_p$. Thus, there are only two solutions with the same norm, but opposite propagation direction:
\begin{equation}
    k_+ = \pm \sqrt{2m \left(E_p - 2\alpha n +\sqrt{(\alpha n)^2+E^2} \right)} \bigg/ \hbar.
\end{equation}
This is obviously different from the case of evanescent bogolons considered in the main text, where the two inverse decay lengths are different.

For comparison, the energy $E=\hbar \omega$ of a bogolon in the evanescent case can be expressed with respect to the inverse decay length $\kappa$:
\begin{equation}
    E^2=\left( -\epsilon_\kappa+\alpha n -E_p\right)\left( -\epsilon_\kappa + 3\alpha n -E_p\right).
\end{equation}
This expression should be compared with Eq.~\eqref{propdisp}.


Finally, we note that the descriptions of the normal region based on the Schr\"odinger equation and on the Bogoliubov-de Gennes equations (BDG) with exactly zero interactions are equivalent. While it may seem that the BDG equations have two solutions with opposite energies, zero interactions mean that the BdG matrix is already diagonal and the negative energy enters the wavefunction only with the minus sign. Thus, these two components simply correspond to two waves propagating in opposite directions with the same energy. This energy is positive when measured from the bottom of the band.

\subsection{Andreev reflection}

In this section, we present explicit expressions for the reflection coefficients for normal and Andreev reflection.

We start by commenting the limit of vanishing lifetime used for the analytical calculations. A finite $\gamma$ reduces both reflection coefficients. Its effect is stronger for the case of large penetration length $1/\kappa_\pm$, discussed in the main text. This occurs especially for $E\to\Delta$. Therefore, the analytical results that we obtain should not be applied for energies close to the edge of the gap and for particularly narrow gaps, $\Delta\to 0$.

As said in the main text, both the wavefunctions and their derivatives have to be continuous on the interface. This imposes:
\begin{widetext}
\begin{equation}
    \begin{array}{r c l}
    \frac{1}{\sqrt{v_1}}[1+r_{N_p}]\begin{pmatrix}1\\0\end{pmatrix}  
    + \frac{r_{A_p}}{\sqrt{v_2}} \begin{pmatrix}0\\1\end{pmatrix}  & = & 
    \eta_+\begin{pmatrix}u_+\\v_+\end{pmatrix}
    + \eta_- \begin{pmatrix}u_-\\v_-\end{pmatrix}, \\
    \frac{ik_1}{\sqrt{v_1}}[1-r_{N_p}]\begin{pmatrix}1\\0\end{pmatrix}  
    - \frac{ik_2r_{A_p}}{\sqrt{v_2}} \begin{pmatrix}0\\1\end{pmatrix}  & = &
    -\kappa_+ \eta_+\begin{pmatrix}u_+\\v_+\end{pmatrix}
    -\kappa_- \eta_- \begin{pmatrix}u_-\\v_-\end{pmatrix}. 
    \end{array}
\end{equation}
\end{widetext}
Matching the wavefunctions and their derivatives at the interface gives an analytical expression for the reflection coefficients:
\begin{equation}\label{coeffs}
   \begin{array}{r c l}
      r_{N_p} & = & \frac{(k_2+i\kappa_-)(k_1-i\kappa_+)u_+v_- - (k_1-i\kappa_-)(k_2+i\kappa_+)u_-v_+}{ (k_2+i\kappa_-)(k_1+i\kappa_+)u_+v_- - (k_1+i\kappa_-)(k_2+i\kappa_+)u_-v_+},\\
    r_{A_p} & = & \frac{2 i k_1 (\kappa_+-\kappa_-) v_- v_+}{ (k_2+i\kappa_-)(k_1+i\kappa_+)u_+v_- - (k_1+i\kappa_-)(k_2+i\kappa_+)u_-v_+} \frac{\sqrt{v_2}}{\sqrt{v_1}}.
   \end{array}
\end{equation}

The precedent expressions can be reformulated to make the phase appear explicitly by considering the group velocities $v_{1,2}=\hbar k_{1,2}/m$ and the relation between the Bogoliubov coefficients $u_+=-v_-e^{2i\phi}$ and $v_+=u_-e^{-2i\phi}$:
\begin{equation}\label{refl-coeffs}
   \begin{array}{r c l}
      r_{N_p} & = & \frac{(k_1-i\kappa_-)(k_2+i\kappa_+)|u_-|^2 + (k_2+i\kappa_-)(k_1-i\kappa_+)|v_-|^2}{ (k_1+i\kappa_-)(k_2+i\kappa_+)|u_-|^2 + (k_2+i\kappa_-)(k_1+i\kappa_+)|v_-|^2},\\
    r_{A_p} & = & \frac{2 i \sqrt{k_1k_2} (\kappa_+-\kappa_-) |u_-||v_-| }{ (k_1+i\kappa_-)(k_2+i\kappa_+)|u_-|^2 + (k_2+i\kappa_-)(k_1+i\kappa_+)|v_-|^2}e^{-2i\phi}.
   \end{array}
\end{equation}

This form makes appear explicitly the role played by $\phi$, and more specifically the change of sign of the Andreev reflection coefficient when $\phi=\pi/2$. The reflection coefficients for holes are given as $r_{N_h}(E)=r_{N_p}(-E)$ and $r_{A_h}=r_{A_p}e^{4i\phi}$.

The dependence on the different energies at stake is here hidden in the complexity of the formulas. However, for  $E\ll E_p$, an approximate expression can be given for both coefficients:
\begin{equation}
    \begin{array}{r c l}
    r_{N_0} & = & \frac{E_p+\sqrt{\left(\alpha n-E_p \right)\left(3\alpha n-E_p \right)}}{E_p-\sqrt{\left(\alpha n-E_p \right)\left(3\alpha n-E_p \right)}+i\sqrt{E_p}(\sqrt{\alpha n-E_p}+\sqrt{3\alpha n-E_p})}\\
    r_{A_0} & = & \frac{e^{-2i\phi}i\sqrt{E_p}(\sqrt{\alpha n-E_p}-\sqrt{3\alpha n-E_p})}{E_p-\sqrt{\left(\alpha n-E_p \right)\left(3\alpha n-E_p \right)}+i\sqrt{E_p}(\sqrt{\alpha n-E_p}+\sqrt{3\alpha n-E_p})}
    \end{array}
\end{equation}

With these expressions, we clearly notice that the three crucial energies to consider are $E_p$, $\alpha n-E_p$ and $3\alpha_n-E_p$. It allows to find the asymptotic values of the reflection coefficients for $\Delta\to 0$.
The maximal value for Andreev reflection coefficients, $|r_A|^2=2/3=2|r_N|^2$, is achieved for $\alpha n\to E_p$ ($\Delta\to 0$). However, we note that these values are beyond the domain of the validity of the theory, since for $E=\Delta=0$ any finite decay $\gamma$ plays a non-negligible role. For realistic $\Delta$, $|r_A|^2<|r_N|^2$.

In the main text, we also consider the case of a SNS junction with a normal region of width $a$. In the derivation of the energy components of a bound state, based on the scattering matrices of the interfaces formed by the reflection coefficients discussed above, we also need the expression of the scattering matrix describing the propagation of the wave in the normal region $S_N$. Regardless of the direction of propagation, this matrix reads:
\begin{equation}
    S_N=\begin{pmatrix}
    e^{ik_1a} && 0 \\
    0 && e^{ik_2a}
    \end{pmatrix}.
\end{equation}
This matrix, together with the ones describing the reflection processes on both interfaces, allow one to find a condition for the existence of a bound state which takes the form of an cancellation of a determinant (see main text). This condition is equivalent to:
\begin{equation}
    |r_A|^2\cos(2\phi)=\Re\left[r_{N_h}^*\left(r_{N_p}e^{ik_-a}-\frac1{r_{N_p}}e^{-ik_+a}\right)\right],
\end{equation}
where $k_\pm = k_1 \pm k_2$, which is more compact and makes appear the reflection coefficients explicitly.

\subsection{Zak phase}
To compute the Zak phase, we use:
\begin{equation}
\Phi_{Zak}  = \int {\left\langle {X}
 \mathrel{\left | {\vphantom {X {i\frac{{\partial X}}{{\partial \varphi }}}}}
 \right. \kern-\nulldelimiterspace}
 {{i\frac{{\partial X}}{{\partial  \phi }}}} \right\rangle } d \phi.
 \end{equation}
 


As in the main text, the wavefunction in the normal (central) region is written as:
\begin{equation}\label{psiN}
    \psi_N=\begin{pmatrix} Ae^{ik_1x} + Be^{-ik_1x}\\ 
    Ce^{ik_2x} + De^{-ik_2x}\end{pmatrix},
\end{equation}
and the associated vectors for each energy component $\pm E$ are written as:
\begin{equation}
    X_{+}=\begin{pmatrix} A' \\ B' \end{pmatrix}~\text{and}~ X_{-}=\begin{pmatrix} C' \\ D' \end{pmatrix},
\end{equation}
where the coefficients are the ones defined in \eqref{psiN} but normalized to one ($|A'|^2+|B'|^2=1$ and $|C'|^2+|D'|^2=1$). The coefficients themselves are computed numerically. Such vectors can indeed be plotted on the Bloch sphere. The vector corresponding to the dominant energy always gives $\Phi_{Zak}=0$ while the other one gives $\Phi_{Zak}= \pi$.

We note that both spinors are constrained to the great circle of the Bloch sphere by the symmetry of the probability density distribution. Indeed, since the problem is completely symmetric with respect to $x=0$, the probability density has to exhibit mirror symmetry with respect to this point. For this, the relative phase between $A$ and $B$ and also between $C$ and $D$ has to be either 0 or $\pi$, which means that the pseudospin can only make a circle through the constant longitude plane (azimuthal angles 0$^\circ$ and 180$^\circ$). If one allows an arbitrary phase between these coefficients, the probability density is shifted and becomes asymmetric. Indeed,
\begin{equation}
    Ae^{ik_1x}+Be^{-ik_1x}=(A+B)\cos k_1 x+i(A-B)\sin k_1 x.
\end{equation}
If the phase difference between $A$ and $B$ is $0$ or $\pi$, we can assume that both are real. In this case, the probability density simply writes
\begin{eqnarray}
   & |Ae^{ik_1x}+Be^{-ik_1x}|^2\nonumber\\
   &=(A+B)^2\cos^2 k_1 x +(A-B)^2\sin^2 k_1 x\nonumber\\
    &=1+2AB\cos^2 k_1 x,
\end{eqnarray}
which is symmetric.
We can now introduce the phase difference between the two coefficients explicitly:
\begin{widetext}
\begin{eqnarray}
     Ae^{ik_1x}+Be^{i\varphi}e^{-ik_1x}&=&e^{i\varphi/2}(Ae^{-i\varphi/2}e^{ik_1x}+Be^{i\varphi/2}e^{-ik_1x}) \nonumber\\
    &=&e^{i\varphi/2}\left[(A+B)\cos (k_1 x-\varphi/2)+i(A-B)\sin (k_1 x-\varphi/2)\right],
\end{eqnarray}
\end{widetext}
which gives an asymmetric probability density
\begin{equation}
    |Ae^{ik_1x}+Be^{i\varphi}e^{-ik_1x}|^2=1+2AB\cos^2 (k_1 x-\varphi/2).
\end{equation}
We conclude that the azimuthal angle on the Bloch sphere $\varphi$ has to be zero, and that the pseudospin has to follow the great circle.

\subsection{Probability current}
 \begin{figure}[h]
    \centering
    \includegraphics[width=0.8\linewidth]{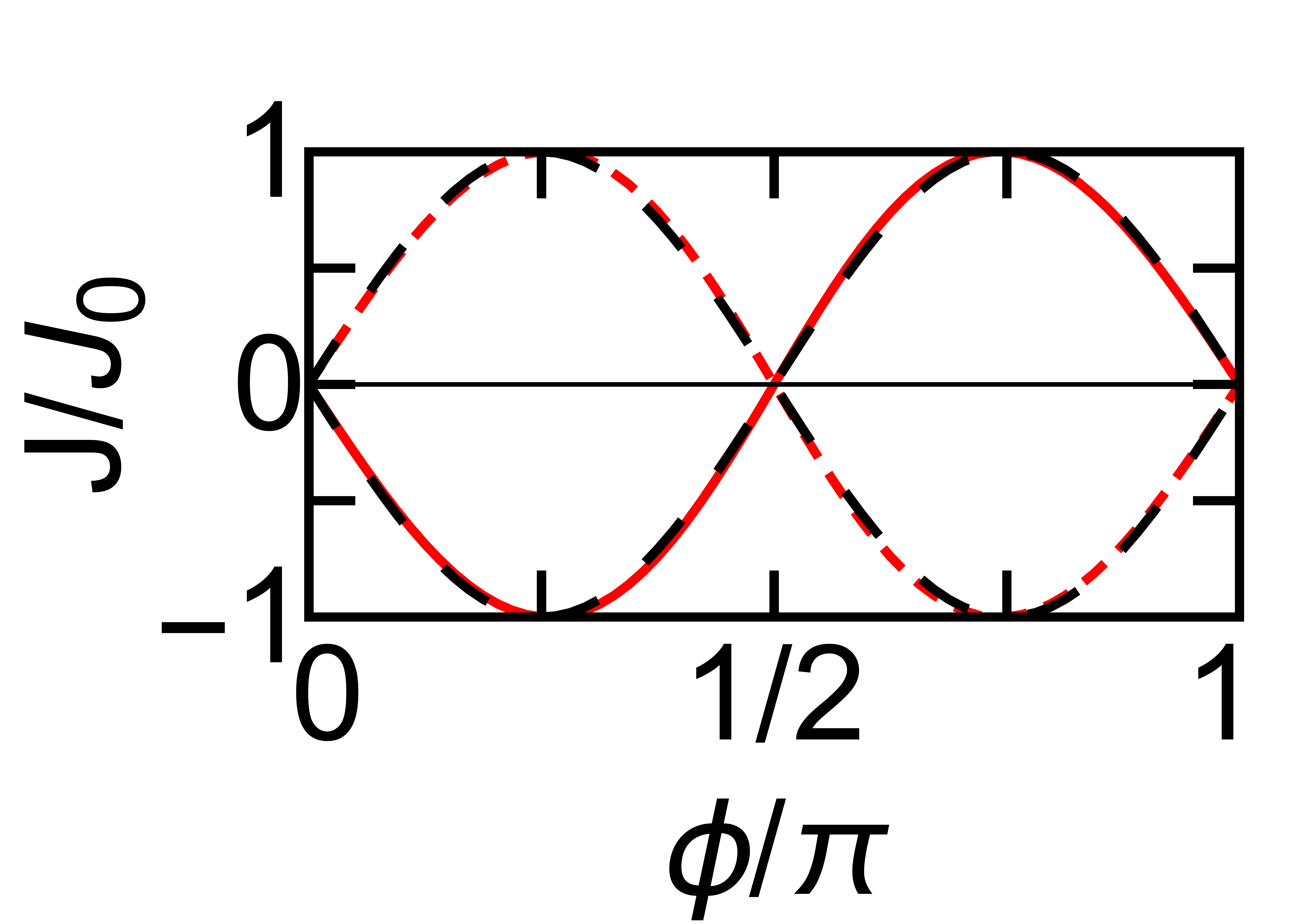}
    \caption{Numerically calculated probability current of the dominant (red straight line) and minority (red dashed line) energy components for a hole-like state. Both follow $\pm \sin(2\phi)$ behaviour (black dashed lines).}
    \label{figSJ}
\end{figure}
The probability current $J_\pm$ of each energy component $\pm E$ can be computed numerically from the expression:
\begin{equation}
    J_+=v_1(|A|^2-|B|^2)~;~J_-=v_2(|C|^2-|D|^2).
\end{equation}
These currents can be plotted with respect to the phase difference  (see Fig.~\ref{figSJ}). Furthermore, by plotting on the same graph $\pm J_0 \sin(2\phi)$, where $J_0$ (which is phase constant) is the maximum absolute value of $J_\pm$, one can notice that there is an excellent match between $J_\pm$ and $\pm J_0 \sin(2\phi)$:
\begin{equation}
    J_\pm \approx \pm J_0 \sin(2\phi).
\end{equation}
This trend can be traced back to the definition of the current on the right interface. The probability current of each energy component measures the exchange between particles at $+E$ and particles at $-E$. Thus, considering the positive energy component for instance, it follows:
\begin{equation}
    J_+ \propto r_{Ah}^R-r_{Ap}^R.
\end{equation}
Regarding the expressions of these coefficients given in Eq.~\eqref{refl-coeffs}, one can deduce that:
\begin{equation}
    J_+ \propto \sin{2\phi}.
\end{equation}
Finally, we retrieve the expression:
\begin{equation}
    J_\pm = \pm J_0' \sin(2\phi),
\end{equation}
where $J_0'$ has a small phase dependence, contrary to $J_0$. This phase dependence comes from the dependence of the energy of the components of a bound state on the phase difference (via the norm of the reflection coefficients).

\subsection{Evanescent states in the normal region}
\begin{figure}[h]
    \centering
    \includegraphics[width=0.9\linewidth]{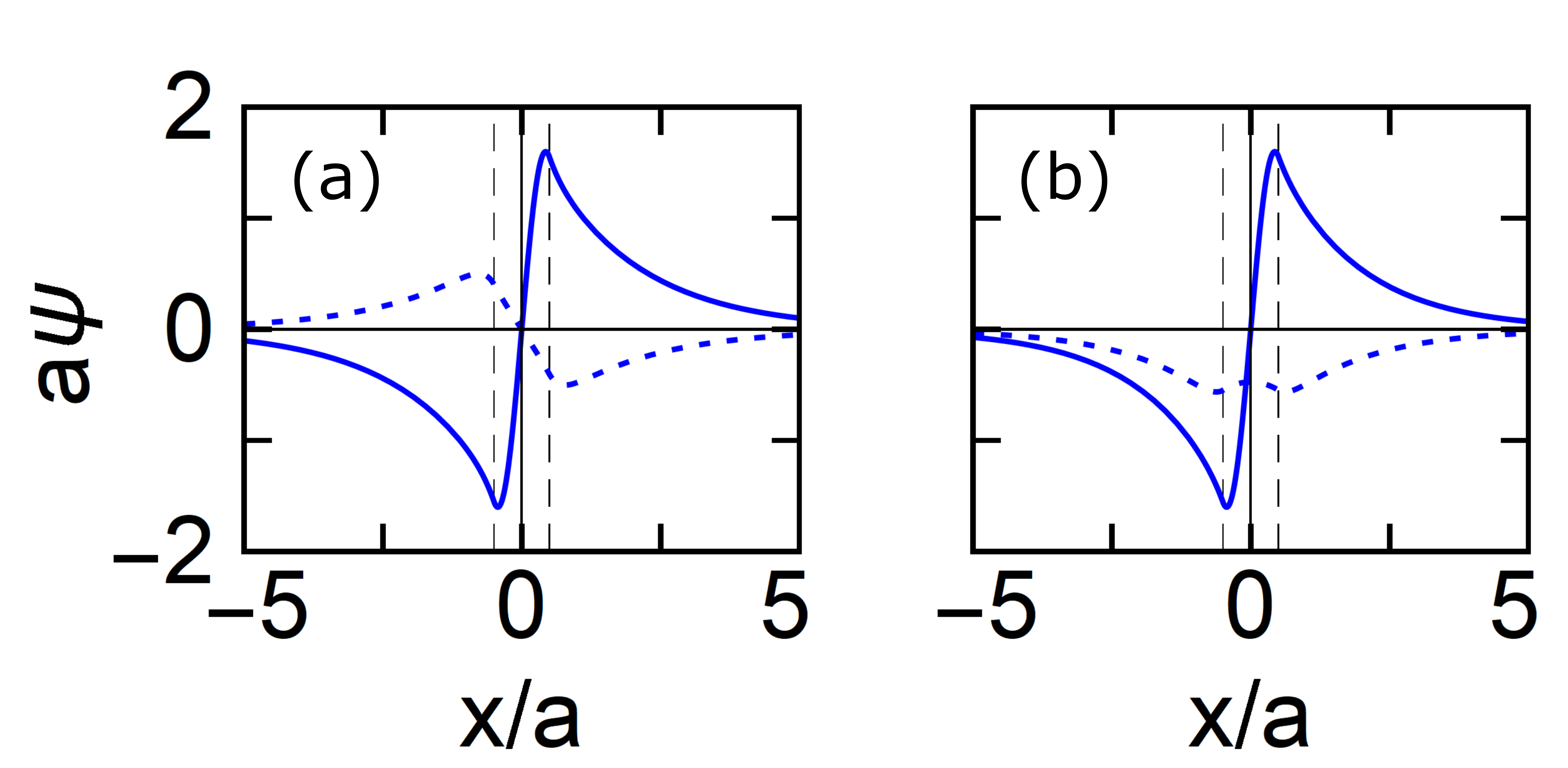}
    \caption{Andreev particle-like bound states for $a=3\,\mu$m and for a phase difference of (a) 0, and (b) $\pi/2$. The dominant (minority) component is shown as solid (dashed) line. Note the symmetry inversion for the minority component.}
    \label{figS1x}
\end{figure}

In the main text, the case $E_p>\Delta$ is considered because it leads to propagative states in the normal region for both positive and negative values of $E$, which is the most interesting case. However, the case with $E_p<\Delta$ is possible as well. Then, for incident particles with energies $E>E_p$, the reflected part at the  energy symmetric with respect to the pump detuning is evanescent. We have solved the reflection problem in this case and obtained a non-zero amplitude of the reflected evanescent wave. Our calculations show that
SNS junctions with this type of states can exist as well. They present the same global behaviour as for the propagative states (see Fig.~\ref{figS1x}(a,b) for the two configurations with a different phase), with the wavefunction in the normal region being a linear combination of hyperbolic functions. However, the bands they form no longer cross. Indeed, the crossing of the bands in the main text occurred when an original state of the quantum well had the same energy as the bogolon image of another state. This is not possible when the original states are propagative and the images are evanescent, since they are always at the opposite sides of zero. Thus, the topologically protected self-amplified interface states discussed in the main text cannot be observed for these bands.
 
\subsection{Non-topological configuration of Andreev bound states}

To confirm that the band crossing and amplification observed in the main text are indeed due to the topology of the bands, we consider an alternative configuration where the bands do not exhibit any inversion of the topology with the variation of the parameter, and thus they anticross instead of crossing each other.
 \begin{figure}[h]
    \centering
    \includegraphics[width=0.8\linewidth]{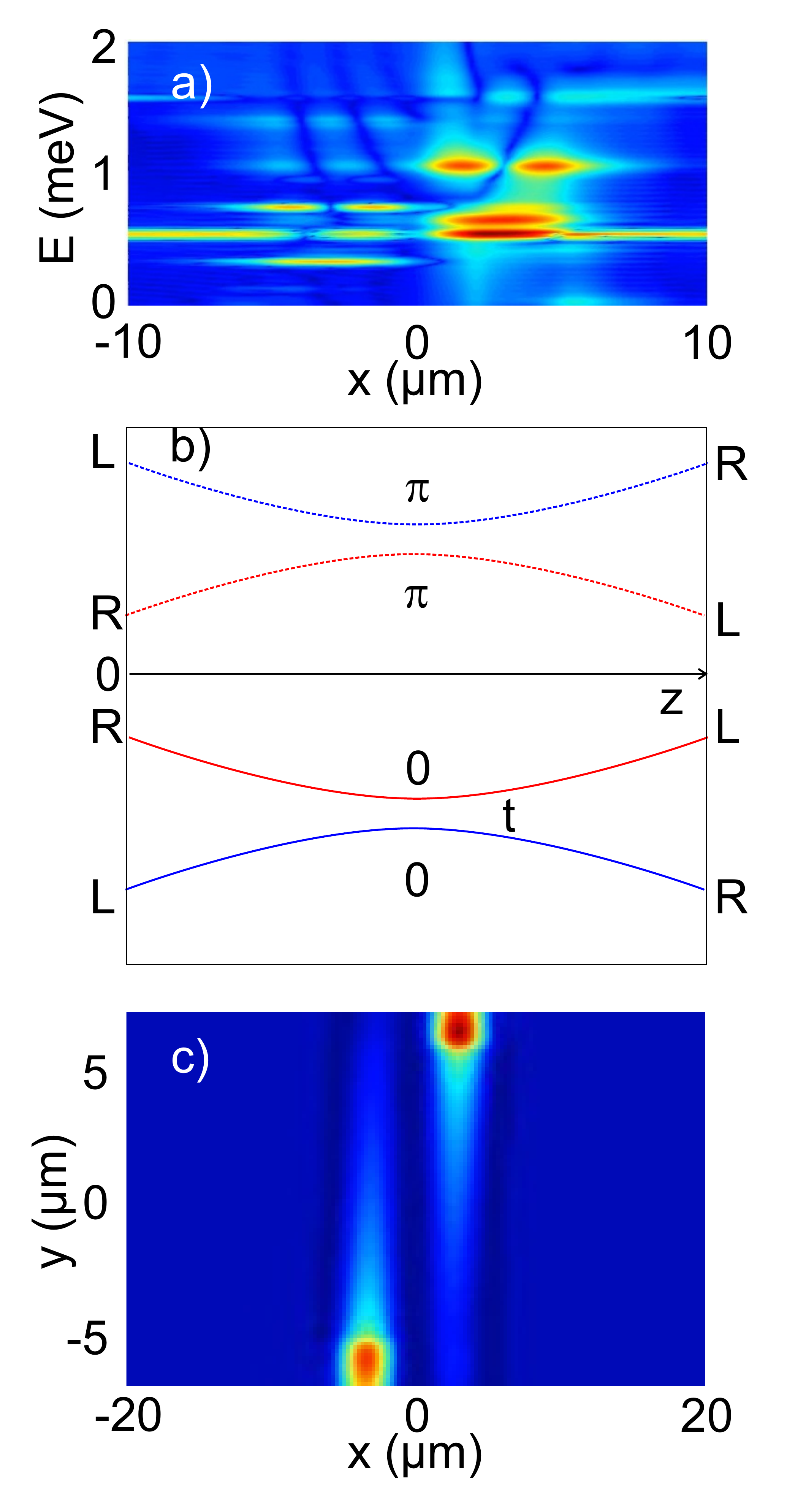}
    \caption{a) The intensity of the Andreev states as a function of position and energy for $U_1 y /l \neq 0$ and $\varphi=\pi/2$. The two original states are strongly detuned, and their bogolon images are detuned as well. b) The variation of the energies of the original states and their images as a function of left-right detuning $U_1 y/l$. The anticrossing is due to the tunneling. The Zak phase indicates that the topology of the bands does not change. c) The real-space image with the detuning varied along $y$: no instability is visible at $y=0$.}
    \label{figS1}
\end{figure}

We introduce an additional potential barrier at $x=0$, splitting the trap into two parts with two distinct trapped states (left- and right-localized) having the same symmetry ($s$ for the lowest state). We also introduce a potential step, responsible for the detuning of these two states. Instead of varying the width of the trap $a$, we vary the height of the step as a function of the second coordinate $y$. The total potential therefore reads
\begin{equation}
    U(x,y)=U_0\exp\left(-\frac{x^2}{2\sigma^2}\right)+U_1 \frac{y}{l} \sign(x)
\end{equation}
where $U_0=2$~meV is the barrier height, $\sigma=0.6$~$\mu$m is its width, $U_1=0.2$~meV is the characteristic step height, and $l=50$~$\mu$m is the characteristic variation length of the step height.

The results of numerical simulations of this configuration are shown in Fig.~\ref{figS1}. Panel (a) presents an example of the calculated spectrum of the Andreev bound states for a particular value of $U_1 y/l=0.15$~meV (the total detuning between the left and right states is $0.3$~meV). The original ($s$-type) states are clearly visible, as well as their bogolon images exhibiting $p$-symmetry due to the laser phase $\phi=\pi/2$. Each of the four visible states belongs to a band (as a function of the synthetic variable $\phi$). The Zak phases of the bands are calculated as in the main text. They are shown in Fig.~\ref{figS1}(b), together with the energies of the band extrema at $\phi=\pi/2$ plotted as a function of the step height. The two lowest bands, formed from the original $s$-symmetric states, have a zero Zak phase. Their symmetry is the same. Thus, when the step height changes sign and the detuning inversion leads to the state inversion (the lowest state changes localization from left to right), the topology of the system does not change. There are no topological reasons for the crossing of the bands, and indeed, it does not occur: their anticrossing is controlled by the tunneling $t$ across the barrier in the center (controlled by its height $U_0$). The same concerns the two upper bands, sharing the same topology (Zak phase $\pi$, different from the two lowest bands). Finally, panel (c) confirms that no amplification due to a band crossing occurs in this case (since the crossing is actually avoided): as the step height changes with $y$, the detuning of the states with respect to the laser changes, and we observe the transfer of maximal intensity from left to right, but no signs of mascroscopically populated oscillating modes are visible.
This confirms that the band crossing and the resulting mode amplification discussed in the main text are indeed of a topological origin.


\end{document}